\documentclass[prd,aps,preprint,amsmath,nofootinbib,amssymb,eqsecnum,showkeys,tightenlines]{revtex4-1}
\pdfoutput=1
\usepackage{slashed}
\usepackage{epsfig,latexsym,cancel,amssymb,amsmath,verbatim,mathrsfs}
\usepackage{color}
\usepackage{graphicx}
\usepackage[para, referable]{threeparttablex}
\usepackage{hyperref}

 \usepackage{mathtools} 
 \usepackage{amssymb}
\definecolor{My_red}{cmyk}{0.00,1.00,1.00,0.20}

\begin{document}
\title{Gamma-ray Signal from $Z_{N\geq 3}$ Dark Matter-Companion Models}
\author{Jun Guo}
\email[E-mail: ]{jguo\_dm@jxnu.edu.cn}
\affiliation{College of Physics and Communication Electronics, Jiangxi Normal University, Nanchang 330022, China}

\author{Zhaofeng Kang}
\email[E-mail: ]{zhaofengkang@gmail.com}
\affiliation{School of physics, Huazhong University of Science and Technology, Wuhan 430074, China}

\author{Ji-Gang Zhao}
\email[E-mail: ]{J.G.Zhao@jxnu.edu.cn}
\affiliation{College of Physics and Communication Electronics, Jiangxi Normal University, Nanchang 330022, China}

\begin{abstract}

In Ref.~\cite{Guo:2021rre}, we proposed to replace the final dark matter (DM) particle in the semi-annihilation mode $\rm DM+DM\to antiDM+Higgs~boson$ with its $Z_{N\geq 3}$ companion, thus reducing DM number density without DM-nucleon scattering.  In this work, we study the indirect detection signals from DM annihilation, the Higgs boson pair with one of them from the companion decay being on- or off- shell, depending on the DM-companion mass splitting. We generate the photon spectrum by using PYTHIA8 and study the properties of the spectrum, to find that the hard part of the spectrum in our model is mainly shaped by the direct Higgs boson and thus does not differ much from that of the conventional semi-annihilation mode. Using the Fermi-LAT data of white dwarfs, we derive the current limit of the DM annihilation cross section for ${\rm DM+DM\to companion^*+Higgs~ boson}$, and for the relatively light DM, it reaches the typical thermal cross section. However, for the TeV scale DM, we have to rely on the Cherenkov Telescope Array, which is able to rule out the whole parameter space except for the coannihilation region.

\end{abstract}
\maketitle

\section{Introduction}

The nature of the putative particle  dark matter (DM)  remains an puzzle in the field of particle physics, and the weakly interacting massive particles (WIMP) have been the most promising candidates due to the WIMP miracle to understand the DM relic density  $\Omega h^2\simeq 0.12$ today.  In general, the WIMP DM is supposed to scatter with nucleons. The accordingly devised DM direct detection experiments such as XENON1T~\cite{XENON:2017lvq} and PandaX~\cite{PandaX:2014mem} have pushed the detection target to the multi-ton-scale level to hunt for the rare  scattering signals between DM and the target atoms, however, no significant signals have been observed yet. Therefore, they merely impose stringent bounds on the parameter space for the typical WIMP DM, whose annihilation cross section is closely tied with the DM-nucleon scattering rate, rendering the WIMP miracle questionable and even the WIMP crisis.

It might be a crisis for the benchmark WIMP DM candidate such as neutralino in supersymmetry and the singlet scalar DM through the Higgs portal. But there are still many WIMPs living a carefree life that can break the general relations between DM-DM annihilation and DM-nucleon scattering. An example is the $Z_{N\geq 3}$ symmetric DM-companion model we proposed before~\cite{Guo:2021rre}, where DM has a companion; moreover, the quantum number assignment admits the semi-annihilation mode: ${\rm DM+DM\to companion^*+Higgs~ boson}$, where the anti-companion will decay back to DM plus some other SM particles, for instance an on or off shell Higgs boson depending on the mass splitting between DM and the companion, $\Delta m$. Remarkably, such kind of semi-annihilation does not give rise to scattering between WIMP DM and nucleon and thus is able to maintain the WIMP miracle and easily evade the direct detection bounds~\footnote{Due to the characterized DM and DM instead of DM anti-DM annihilation mode, such models provide a good opportunity to seed matter asymmetry~\cite{Chen:2024arl}.}. Then, we need to search for this type of WIMP DM by indirectly detect its Higgs boson signals from DM annihilation today.

The best way to look for these products in the sky is to collect their gamma rays injected into the cosmic ray. In this work, based on the simplest model with two identical (except for mass) complex scalar fields  with respect to $Z_3$, we simulated the resulting gamma spectra for different DM-companion mass splitting $\Delta m$, to find that for most interesting cases, the contribution from the direct Higgs boson plays the dominant role while the secondary Higgs boson is relatively soft. As a consequence, our signals, in particular for a sufficiently small $\Delta m$, do not differ much from the conventional $Z_3$ symmetric DM model~ \cite{DEramo:2010keq,Belanger:2012vp,Ko:2014nha,Guo:2015lxa,Guo:2023kqt,DiazSaez:2022nhp,Liu:2023kil}, which produces the single Higgs boson signal  ${\rm DM+DM\to DM^*+Higgs~ boson}$. Therefor, our study also apply to these models; see some variants with more complicated model extensions~\cite{Queiroz:2019acr,Toma:2021vlw}.

Then, we derive the indirect detection bound on this model using the Fermi-LAT observation, which covers the energy range between 20 MeV and 300 GeV and has been observing the gamma-ray sky for 15 years, drawing the most detailed map to date. Many works have made a deep analysis of the data, focusing on a wide variety of targets. For instance, Ref.~\cite{Li:2013qya,Karwin:2020tjw} tried to search for the DM annihilation signals in the direction of the Andromeda galaxy (M31), while Ref.~\cite{Fermi-LAT:2015sau,Karwin:2016tsw,Fermi-LAT:2017opo,DiMauro:2021raz} studied the data from the Galactic Center and founded hints for gamma-ray excess, but it is not a significant DM signal. In particular, Ref.~\cite{Fermi-LAT:2010cni,Fermi-LAT:2011vow,Geringer-Sameth:2014qqa,Fermi-LAT:2015att,Fermi-LAT:2015ycq,Fermi-LAT:2016uux,Calore:2018sdx,Hoof:2018hyn,DiMauro:2021qcf,McDaniel:2023bju} analyzed the Milky Way dwarf spheroidal satellite galaxies (dSphs) using the LAT data, and placed the most stringent constraint on signals from DM activities. The dSphs of the MilkyWay are the most DM dominated celestial bodies and lack of non-thermal astrophysical processes, so they are the most prominent targets for searching for DM decay or annihilation. In this work, we will use the dSphs of the MilkyWay to obtain the upper bound on the cross section for ${\rm DM+DM\to companion^*+Higgs~ boson}$. We find that the current bound reaches the typical thermal WIMP DM cross section for DM below the TeV scale. Whereas for the even heavier DM above the TeV scale, sensitivity improvement of order of magnitude is required, and the Cherenkov Telescope Array (CTA)~\cite{CTAConsortium:2017dvg} offers an opportunity to cover this parameter space.

The paper is organized as follows, in Section~\ref{sec:model}, we will briefly introduce our $Z_3$ symmetric DM-companion model. In Section~\ref{sec:spectrum}, we will study the spectral properties of dark matter annihilation in our model. In Section~\ref{sec:data}, we will first present the data that we will use, and then derive the upper limit of the DM annihilation cross section. Finally, we will give our conclusions in Section~\ref{sec:conclusion}.

\section{$Z_3$ symmetric DM models: from Higgs to companion portal }
\label{sec:model}

The $Z_3$ symmetric model is characterized by the presence of semi-annihilation mode. The simplest model involves a complex scalar $S_1$ transforms under $Z_3$ as $S_1\rightarrow e^{ik 2\pi/3}S_1$ with $k=1$ or $2$, and the resulting Lagrangian is given by
\begin{align}
 -{\cal L}_{Z_3}\supset & m_{1}^2S_1 S_1^* + \lambda_{1h}|S_1|^2|\Phi|^2+\left( \frac{A_1 S_1^3}{3} + c.c\right),\label{Model}
\end{align}
the real scalar part component of $\Phi$ is the Higgs boson $H$, and the cubic term is new compared to the usual $Z_2$ symmetric Higgs portal dark matter model and gives rise to the characterized semi-annihilation mode $SS\to S^*H$, which, through a properly large $A_1$, provides a way to attenuate the stringent exclusion to the usual $Z_2$ Higgs portal dark matter from direct matter detection. But this attenuation is not remarkable~\cite{Athron:2018ipf} since the effective dark matter annihilation rate still requires a sizable Higgs portal coupling $\lambda_{1h}$. 

A complete attenuation is available via the mechanism proposed by us in Ref.~\cite{Guo:2021rre}. It considers at least one more dark species charged under $Z_N$ (dubbed as companion), and then dark matter can gain correct relic density via the semi-annihilation into the companion; the Higgs portal that generates DM-nucleon interaction can be removed by hand or by symmetry. The mechanism can be realized simply by extending the model specified by Eq.~(\ref{Model}) with one more scalar $S_2$, a copy of $S_1$ in quantum number~\footnote{
The model has an updated version if dark matter is instead a Dirac fermion $\Psi$, which automatically does not have Higgs portal interaction at all. However, such models will have very different DM indirect detection signals than the photon signals discussed in the current work, and we leave more details in the discussion.}.  The resulting most general Lagrangian is tedious, and it is unnecessary to list them all. So, we just present the relevant terms  in the mass basis of $S_{1,2}$, 
\begin{align}\label{Z3L}
 -{\cal L}_{Z_3}\supset m_{S_1}^2S_1 S_1^* + m_{S_2}^2 S_2S_2^* + 
 \left( \frac{A_1 }{3}S_1^3 +\mu_{12}S_1S_2^*H+c.c.\right).
\end{align}
Other terms, in particular $|S_1|^2|\Phi|^2$, are assumed to be small and irrelevant. The key feature is that the dark matter candidate $S_1$ does not have the conventional Higgs portal term as in Eq.~(\ref{Model}). However, as long as the masses lying in the window
\begin{align}\label{window}
 m_{S_2}>m_{S_1} >(m_{S_2}+m_H)/2>m_{S_2}/2,
\end{align}
terms in the bracket will give rise to two-body semi-annihilation mode for a pair of dark matter particles $S_1$: 
\begin{align}
    S_1S_1\to S^*_2(\to S^*_1+H~{\rm or }~H^{off}) +H,
\end{align}
with $H^{off}$ denoting for the off-shell Higgs boson, which is the situation if the masses of dark matter and companion are sufficiently close to each other, $\Delta m\equiv m_{S_2}-m_{S_1}<m_H=125.6$ GeV; it is a typical case in the relatively light dark matter region. 

In particular, in the highly degenerate region ($\Delta m/m_{S_1}\lesssim 20\%$), the model actually resembles the previous model in the sense of indirect detection. This case corresponds to the coannihilation region analyzed in Ref.~\cite{Guo:2021rre}, which may lead to the reduction of $\langle\sigma v\rangle_{S_1S_1\to S_2^*H}$ given a large $\langle\sigma v\rangle_{S_2S_2^*\to HH}$.

\section{DM annihilation and photon spectrum}
\label{sec:spectrum}

In this section we analyze the photon spectra from the Higgs bosons produced in different annihilation modes:   ${\rm DM+DM\to DM^*+Higgs~ boson}$ (from the conventional $Z_3$ Higgs portal model), ${\rm DM+DM\to companion^*+Higgs~ boson}$ (from the $Z_3$ DM-companion model) and ${\rm DM+DM\to Higgs~ boson~pair}$ (from the conventional $Z_2$ Higgs portal model). The photon spectra of the first and third modes are determined solely by DM mass and their features are clear, while the second case involves one more parameter and is more complicated. Hence, our subsequent analysis will focus on the second mode.

For the model we studied, the DM annihilation mode is $S_1 S_1 \rightarrow H S_2^*$, and the followed decay of $S^*_2$ depends on the mass splitting $\Delta m$. When $\Delta m> m_{H}$, $S_2^*$ decays into DM plus an on-shell Higgs boson  (the secondary Higgs boson), while in the case of $\Delta m\leq m_{H}$, the on-shell Higgs boson becomes off shell, denoted as  $H^{off}$. For both cases, DM annihilation today yields a characteristic photon spectrum. To get these spectra, we take the following procedure: 
\begin{enumerate}
    \item First, we generate the model file using FeynRules ~\cite{Alloul:2013bka}.
    \item  Then, we generate the process $S_1 S_1 \rightarrow H S_2^*$ using MadGraph~\cite{Alwall:2014hca} without hardronization and detector simulation.
    \item  Next, we pass the events to PYTHIA8~\cite{Sjostrand:2014zea} to implement hardronization, and finally obtain the DM annihilation photon spectrum. 
\end{enumerate}
In the rest part of this section, we will discuss the two cases separately, and we will also compare them with those from other two cases.

\subsection{Photon spectrum with an off-shell Higgs boson from companion decay}

\begin{figure}[htbp]
\centering
\includegraphics[width=0.43\textwidth]{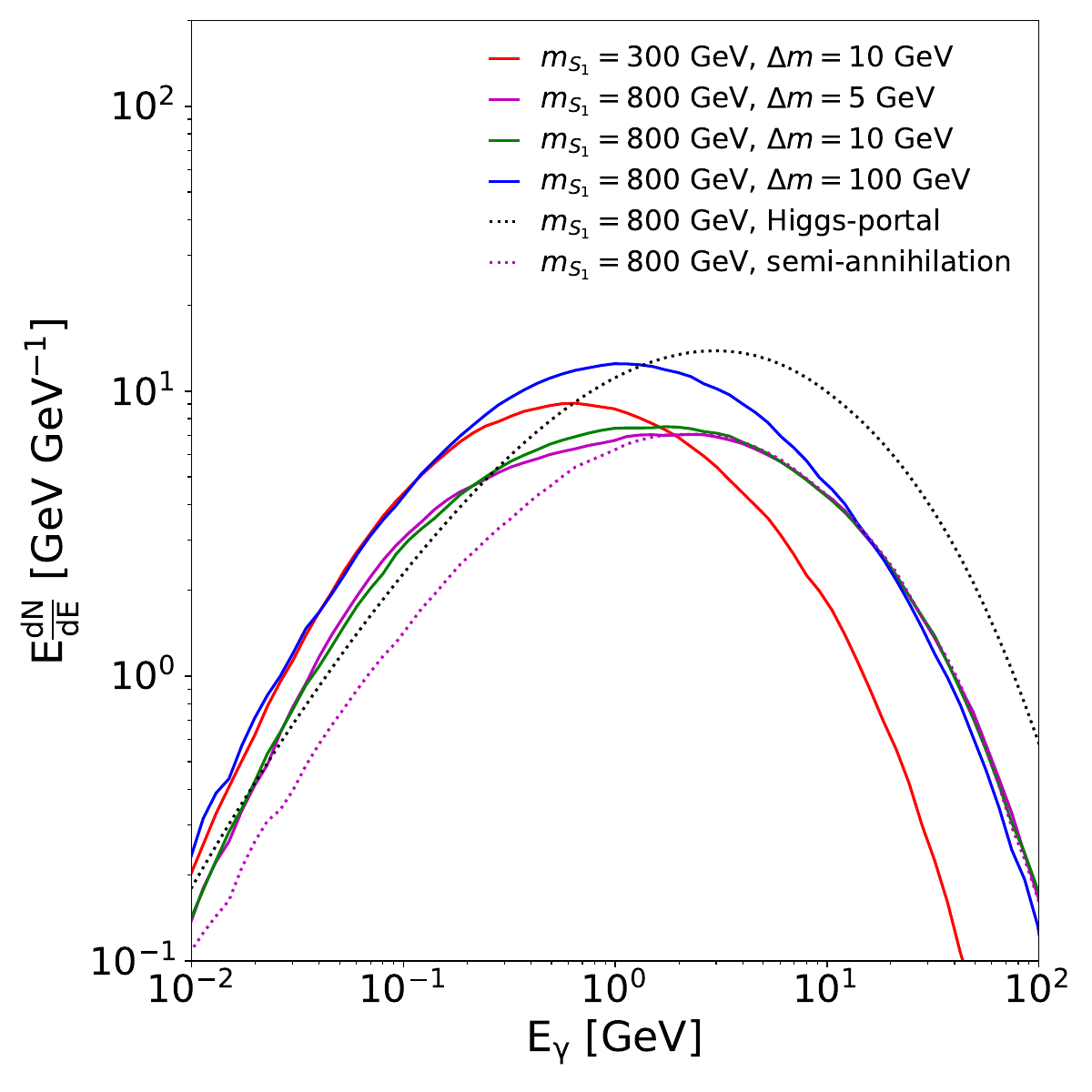}
\includegraphics[width=0.43\textwidth]{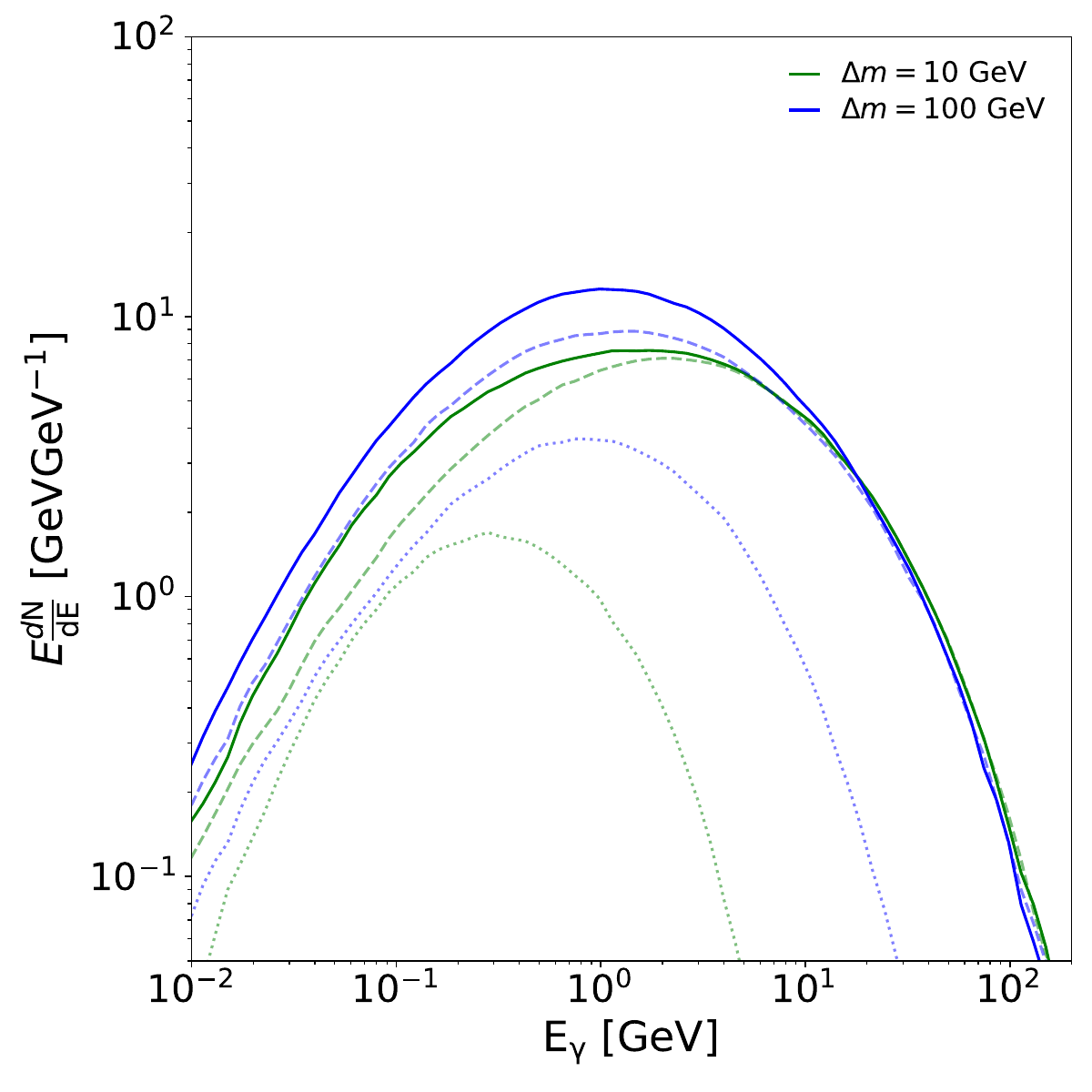}
\caption{Left: Photon spectra for different dark matter mass $m_{S_1}$ and DM-companion mass splittings $\Delta m$ for the $HS_2^*(\to H^{off}+\rm DM)$ mode. To make a comparison, the photon spectra from the $HS$ mode (dotted pink) and $HH$ mode  (dotted black) are shown. Right: two samples of the components of the photon spectrum for the $HS_2^*(\to H^{off}+\rm DM)$ mode with $\Delta m=10$ GeV (green) and $\Delta m=100$ GeV (blue), taking $m_{S_1}=800$ GeV, where the total spectrum (solid) is a sum of the spectra from the direct Higgs boson (dashed) and secondary Higgs boson (dotted). }
\label{off_shell_spectrum}
\end{figure}

Both DM mass $m_{S_1}$ and DM-companion mass splitting $\Delta m$ can affect the shape of the photon spectrum from DM annihilation. When $m_{S_1}$ becomes larger, both the annihilation products, the DM companion $S_2^*$ and the Higgs boson $H$, would be more boosted and then give rise to more high energy photons. This is clearly seen in the left panel of Fig.~\ref{off_shell_spectrum}, the photon spectrum for 300 GeV DM is much softer than that for the 800 GeV DM. 

The photon spectrum is affected by the DM-companion mass splitting from two aspects, the primary (accessible) decay channels of $H^{off}$ and as well the boost of its decay products. In this work, we consider two levels of splitting and the corresponding dominant decay modes of $S_2^*$: 1) A relatively large splitting with $m_H>\Delta m\geq 2 m_{b}$, and then the dominant decay channel of $S^*_2$ is $S^*_2\rightarrow S^*_1 b\bar{b}$; 2) A relatively small splitting with $2m_\tau<\Delta m < 2 m_b$,  and then $S^*_1\tau\bar{\tau}$ dominates the decay products of $S^*_2$, as show on the Fig.~\ref{off_shell_spectrum} left panel, we compare the cases of  $\Delta m = 5$ GeV and $\Delta m = 10$ GeV for the DM mass of 800 GeV, and find that the difference between them is tiny. Even smaller splitting is not considered since it means a significant fine-tuning of the DM-companion mass spectrum. The impact on boosting the fermion pair can be seen from the energy of $S_2^*$ (in the center mass frame) produced in dark matter annihilation:
\begin{align}
    E_{ S^*_2} \simeq m_{  S_1} + \frac{(m_{  S_1} + \Delta m)^2 - m_{ H}^2}{4m_{  S_1}}.
\label{Eq:boost}
\end{align}
Therefore, for a fixed DM mass, the resulting fermion pair becomes more energetic as $\Delta m$ increases. Its impact on the final photon spectrum is complicated, since the fermion pair comes from $S^*_2$ three-body decay, and thus in the following we give numerical samples, shown on the left panel of Fig.~\ref{off_shell_spectrum}. 

Our samples are for a given DM mass of 800 GeV, with two cases of mass splitting: the smaller $\Delta m=10 $ GeV and the larger $\Delta m=100 $ GeV. It is found that the latter generates a slightly softer photon spectrum, which is mainly ascribed to the soften of the direct Higgs boson $H$. To demonstrate this, we show the components of the photon spectrum in the right panel of Fig.~\ref{off_shell_spectrum}. From it one can see that, in each case, the component from the direct Higgs boson (dashed line) always exceeds that from $S_2^*$ or the off-shell Higgs boson  (dotted line), over the whole spectrum; in particular in the energy region above around 10 GeV, the latter is almost negligible. Moreover, compared to $\Delta m=10 $ GeV, in the case of $\Delta m=100 $ GeV, the $H$ produces fewer photons starting from $E_\gamma \gtrsim 5 $ GeV. Although in the latter case, $H ^ {off} $ generates more photons starting from $E_{\gamma} \gtrsim 1 $ GeV,  this secondary component only gives a mild modification to the spectrum generated by $H$~\footnote{One can understand this along this line: a larger $\Delta m$ means a larger mass of DM companion $S_2^*$, which then takes away more kinetic energy according to Eq.~(\ref{Eq:boost}) , but at the same time, the direct Higgs boson, the main source of photons, is less boosted, so the harder photons from its cascade decay is reduced. }. 

Therefore, in this sense, the indirect detection signals from the ${\rm DM}+HH^{off}$ mode resembles that from the ${\rm DM}+H$ mode contaning a single direct Higgs boson. This is in accordance with the expectation that the $Z_3$ symmetric DM-companion model reduces to the conventional $Z_3$ Higgs portal model when $\Delta m\ll m_{S_1}$; explicitly, for the given DM mass 800 GeV, we compare the photon spectra from the former with that from the latter (dotted pink) in the left panel of Fig.~\ref{off_shell_spectrum}. We find that, for a sufficiently small mass splitting, they coincide in the hard part and only show a mild difference in the soft part below the GeV scale. Hence, the constraint on $\langle \sigma v\rangle_{S_1 S_1 \rightarrow H S_2^*}$ at a given dark matter mass made in this limit can be well applied to $\langle \sigma v\rangle_{S_1 S_1 \rightarrow H S_1^*}$ from the conventional $Z_3$ Higgs portal model.

The photon spectrum from direct Higgs boson pair mode is also shown in the left panel of Fig.~\ref{off_shell_spectrum}, and one can see that it is obviously more energetic than other cases: roughly speaking, the hardness of its spectrum is roughly twice than others in the relatively high energy region.  The reason is not difficult to understand. For this DM annihilation mode, all the energy of the annihilating DM pair is transferred to the Higgs boson pair, whereas in the semi-annihilation modes, both in the conventional $Z_3$ Higgs portal model or its DM-companion variant, the DM particle contained in the final state always takes away a substantial portion of the initial DM energy. This argument means that the photon spectrum corresponding to the combination of a direct Higgs boson and secondary on-shell Higgs boson discussed in the following will not give a much significant difference.

\subsection{Photon spectrum with an on-shell Higg boson from companion decay}

In the case of $\Delta m>m_{H}$, in addition to the direct Higgs boson, DM direct annihilation produces another on-shell Higgs boson, the secondary Higgs boson from $S_2^*$ two-body decay. Unlike the previous case, here the mass splitting $\Delta m$ can only affect the decay spectrum in one way, the kinetic boost of the Higgs bosons. 

\begin{figure}[htbp]
\centering
\includegraphics[width=0.4\textwidth]{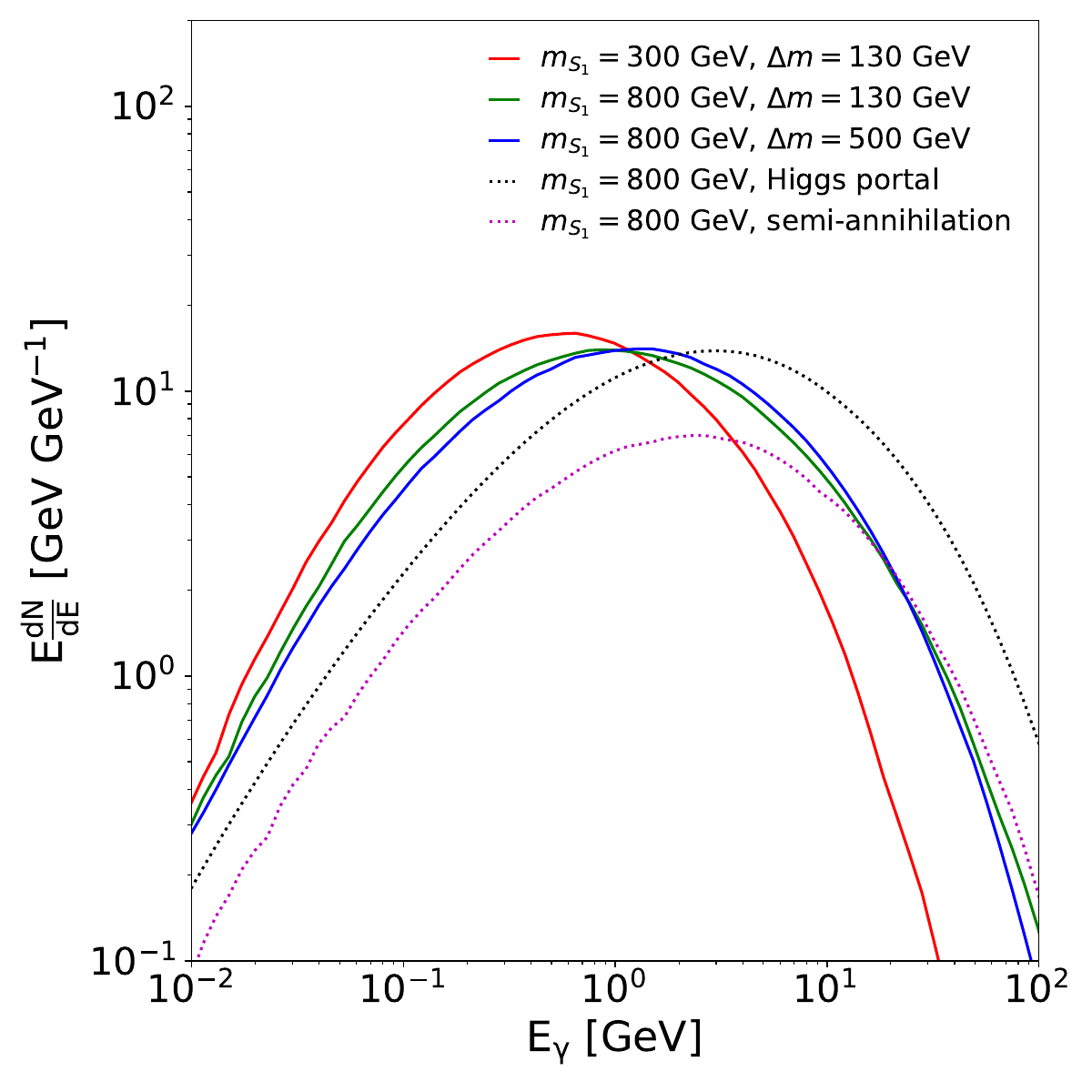}
\includegraphics[width=0.4\textwidth]{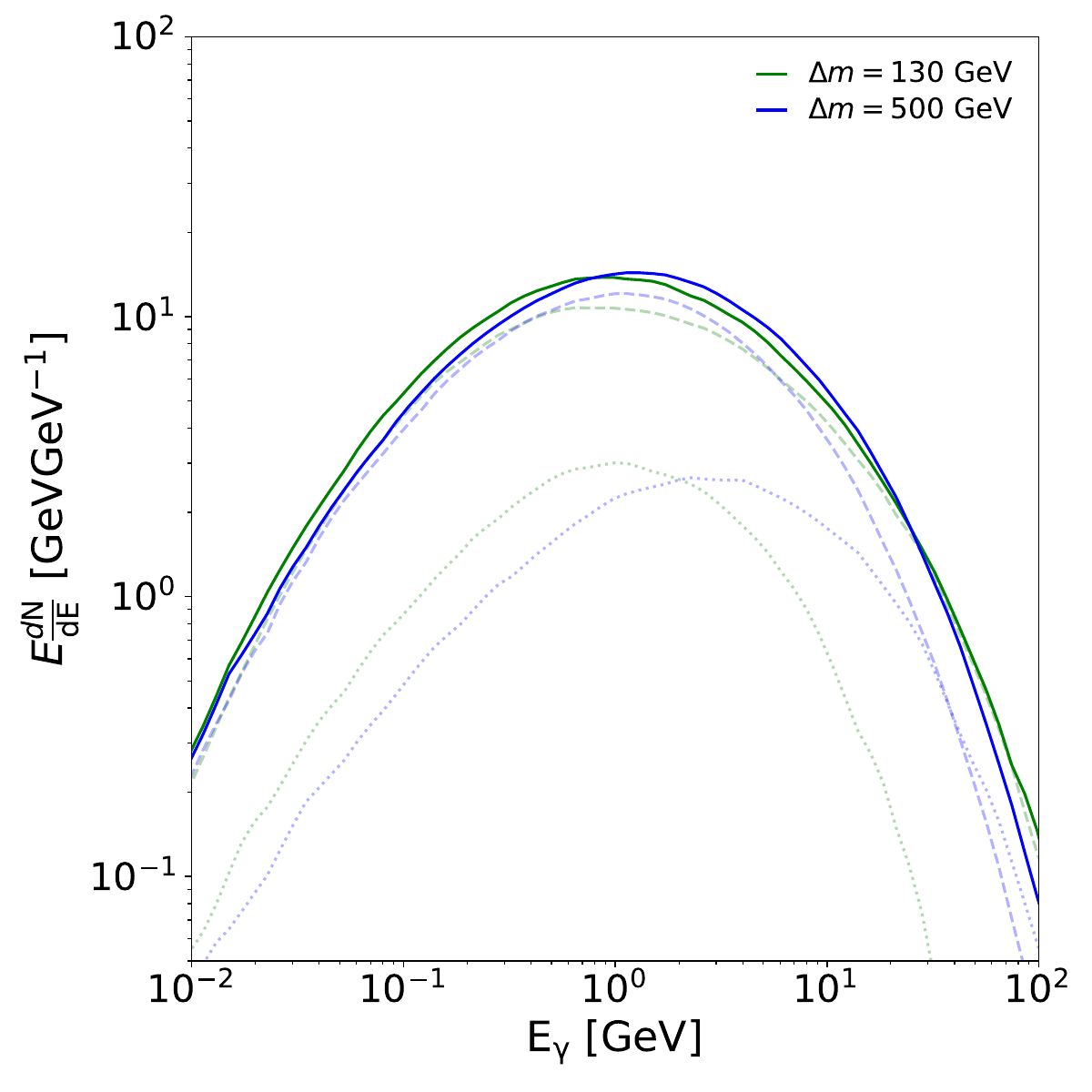}
\caption{Left: Photon spectra for different DM mass and DM-companion mass splittings for the $HS_2^*(\to H+\rm DM)$ mode. Right: two samples of the components of the spectrum with $\Delta m=130$ GeV (green) and $\Delta m=500$ GeV (blue) with $m_{S_1}=800$ GeV, where the total spectrum (solid) is a sum of the spectra from the direct  (dashed) and secondary (dotted) Higgs bosons, respectively.}
\label{on_shell_spectrum}
\end{figure}

The shape of the photon spectrum is again controlled by $m_{S_1}$ and $\Delta m$. From the left panel of Fig.~\ref{on_shell_spectrum}, it is ready to find that the shape mainly depends on the DM mass. As for the mass splitting, its impact is not so significant compared to the previous case containing $H^{off}$. To understand the reason, in the right panel of Fig.~\ref{on_shell_spectrum}, we show the photon spectra for $\Delta m=130$ GeV and $\Delta m=500$ GeV, respectively, both having DM mass $m_{S_1}=800$ GeV. We can see that, the photon spectrum again is almost dominated by the component from the direct Higgs boson, however, for the large mass splitting, when $E_\gamma\gtrsim {\cal O}(40)$ GeV, the component from the secondary Higgs boson begins to be comparable and eventually dominated at the high-energy tail region. 

Once again, we compare the photon spectra in the ${\rm DM}+2H$ mode with other spectra in Fig.~\ref{on_shell_spectrum}, taking a 800 GeV dark matter. It is seen that, as previously commented, they are indeed similar to the shape of the DM$+H$ mode in the relatively higher energy range, and it is even harder for $E_\gamma \gtrsim {\cal O}(10)$ GeV, by virtue of its more boosted direct Higgs boson. Therefore, the real reason is traced back to why the secondary Higgs boson again plays a minor role. As mentioned before, it is due to the reason that most of the energy of $S_2^*$ is taken away by DM, so the secondary Higgs boson is relatively soft. Only when the mass splitting is large enough, $\Delta m\gg m_H$, can it be significantly accelerated, thus hardening its component in the photon spectrum. At the same time, the direct Higgs boson becomes softer by increasing $\Delta m$, which is seen from the expression of its energy, 
\begin{align}
    E_{H} \simeq m_{S_1} - \frac{(m_{S_1} + \Delta m)^2 - m_{H}^2}{4m_{S_1}}.
\label{Eq:boost}
\end{align}
It is substantially reduced when $\Delta m$ becomes comparable to $m_{S_1}$.



\section{Indirect Detection}

After analyzing the characteristic photon spectrum from the Higgs boson ``pair", injected by DM semi-annihilation into the cosmic ray, the main subject of this section is to apply the FERMI-LAT data to make constraint on such kind of DM models. But the prospect at CTA will also be briefly studied.

\subsection{Injection, propogation and observation}

The indirect detection experiments such as FERMI-LAT observe the gamma-ray flux signal in a particular region of the sky. The expected signal from the DM hypothesis of the observed sky map region in the energy range $[E_{\rm min}, E_{\rm max}]$ is factorizied as:
\begin{align} \label{gamma_flux}
\phi(\Delta \Omega; E_{\rm min}, E_{\rm max}) = \frac{1}{4\pi}\frac{\langle \sigma v\rangle}{2m_{\rm DM}^2}\int_{E_{\rm min}}^{E_{\rm max}}\frac{dN_\gamma}{dE_\gamma}\times \rm J,
\end{align}
where $\langle \sigma v\rangle$ is the thermally averaged annihilation cross section of DM times the relative velocity, and the $\frac{dN_\gamma}{dE_\gamma}$ is the DM annihilation photon spectrum of a particular DM model. The $\rm J$-factor is the integration of DM density square along the line of sight $s$ and solid angle $\Delta\Omega$, given as:
\begin{align}
{\rm J} = \int_{\Delta\Omega} \int_{l.o.s}d\Omega ds\rho^2(r(s, l, b)),
\end{align}

where $r$ is the radial distance from the Galactic Center, 
\begin{align}
r = \sqrt{s^2 + R^2 - 2sR \cos{l}\cos{b}},
\end{align}
where $R$ is the distance from the Sun to the Galactic Center, s is the distance to the Earth and $(l, b)$ is the Galactic coordinates. $\rho(r)$ is the DM density profile, in this study, we are setting the limitation bound by using the dSphs data, and it is found that the $\rm J$-factor is fairly insensitive to it~\cite{Martinez:2009jh, Strigari:2012acq}. Therefore, we ignore the effect of DM density profile to the indirect detection signal, and the J-factor we use could be found in ~\cite{Fermi-LAT:2015att, McDaniel:2023bju}.

\subsection{Fermi-LAT dSphs Data Analysis and Limitations}\label{sec:data}

We will adopt the available Fermi-LAT dSphs gamma-ray data for the analysis in this work. In ~\cite{Fermi-LAT:2015att}, the authors analyzed the 6 years of Fermi-LAT dSphs data  processed with $\rm pass$ 8. They analyzed 15 dSphs with measured J-factor, and  provided the spectrum energy distribution (SED) for each dSph. For each SED, the energy is divided into 24 energy bins, and a likelihood table of the photon energy flux values is provided in each bin. Recently, authors of Ref.~\cite{McDaniel:2023bju} have generously provided the latest 14.3 years of LAT data. They analyzed  50 dSphs, and divided them into three sets: 1) a measured set which contains 30 dSphs with measured J-factors; 2) the benchmark set which contains 42 dSphs, some of which only have estimated J-factors; 3) the inclusive set which contains all the 50 dSphs. In their analysis, all the dSphs gamma-ray data is selected from the $\rm P8R3\_SOURCE\_V3$ class, with the SEDs ranging from 500 MeV to 1 TeV, also divided into 24 bins. 

We calculate the upper limit of Fermi-LAT dSphs data to our model by performing a joint likelihood analysis, which is  $\Pi_i\mathcal{L}_i$, the product of the likelihood of dSphs of interest. For each such dSph, the likelihood function for DM hypothesis is a product of the combined likelihood of dSph photon flux and the likelihood of dSph J-factor,
\begin{align}
\mathcal{L}_i(m_{\rm DM}, \langle \sigma v\rangle|\mathcal{D}_i) = \mathcal{L}_i(m_{\rm DM}, \langle \sigma v\rangle|\mathcal{D}_i)\mathcal{L}_J(J_i|J_{obs, i}, \sigma_i).    
\end{align}
The former is the production over the likelihood of each energy bin $E_n$,  $\prod_{n}\mathcal{L}_{E_n}(\frac{d\phi}{dE}, E_n)$, where $\frac{d\phi}{dE}$ is the DM photon-flux,  a function of $m_{\rm DM}$, $\langle\sigma v\rangle$ and the J-factor. The likelihood function of dSph J-factor takes the form of 
\begin{align}
\mathcal{L}_J(J_i|J_{obs, i}, \sigma_i) = \frac{1}{\ln{10}J_{obs, i}\sqrt{2\pi}\sigma_i}\exp(-\left(\log_{10}(J_i)-\log_{10}(J_{obs, i})\right)^2/2\sigma_i^2).
\end{align}
In our analysis, we will use the benchmark set to analyze the 14-years data provided in~\cite{McDaniel:2023bju}, with the  J-factor value $J_{obs, i}$ and the uncertainty $\sigma_i$ shown in Table.~\ref{tableJfactor14}. For the 6-years data analysis, we use the 15 dSphs provided in~\cite{Fermi-LAT:2015att}, with the  J-factor value $J_{obs, i}$ and the uncertainty $\sigma_i$ given in Table.~\ref{tableJfactor6}.


To determine the best-fit model under the maximum likelihood estimation, we use the test statistic (TS) which is defined as~\cite{Cowan:2010js}
\begin{align}
\rm{TS} = -2\ln{\frac{\mathcal{L}_{max}}{\mathcal{L}(m_{DM}, \langle \sigma v\rangle|\mathcal{D})}}
\end{align}
where $\mathcal{L}_{\rm max}$ is the maximum likelihood.

\begin{table}[h!]
\centering
\begin{threeparttable}
\caption{ \label{tab:dsphs} Milky way dSphs' properties for \cite{Fermi-LAT:2015att}}
\begin{tabular}{ |l |r |r| c |c|}
\hline
  Name                      & $\ell$ & $b$ & Dist. & $\log_{10}({\ensuremath{J_{\textrm{obs}}}})$ \\
  & (deg) & (deg) & (kpc) & {\small ($\log_{10}(\text{GeV}^2 \text{cm}^{-5})$)}\\
  \hline
  Bootes I                  & 358.1  & 69.6   & 66     & $18.8 \pm 0.22$  \\
  Canes Ven. II         & 113.6  & 82.7   & 160    & $17.9 \pm 0.25$ \\
  Carina                    & 260.1  & $-$22.2  & 105    & $18.1 \pm 0.23$  \\
  Coma Ber.            & 241.9  & 83.6   & 44     & $19.0 \pm 0.25$ \\
  Draco                     & 86.4   & 34.7   & 76     & $18.8 \pm 0.16$  \\
  Fornax                    & 237.1  & $-$65.7  & 147    & $18.2 \pm 0.21$  \\
  Hercules                  & 28.7   & 36.9   & 132    & $18.1 \pm 0.25$ \\
  Leo II                    & 220.2  & 67.2   & 233    & $17.6 \pm 0.18$  \\
  Leo IV                    & 265.4  & 56.5   & 154    & $17.9 \pm 0.28$  \\
  Sculptor                  & 287.5  & $-$83.2  & 86     & $18.6 \pm 0.18$  \\
  Segue 1                   & 220.5  & 50.4   & 23     & $19.5 \pm 0.29$  \\
  Sextans                   & 243.5  & 42.3   & 86     & $18.4 \pm 0.27$  \\
  Ursa Maj. II             & 152.5  & 37.4   & 32     & $19.3 \pm 0.28$  \\
  Ursa Minor                & 105.0  & 44.8   & 76     & $18.8 \pm 0.19$ \\
  Willman 1                 & 158.6  & 56.8   & 38     & $19.1 \pm 0.31$  \\
  \hline
\end{tabular}
\label{tableJfactor6}
\end{threeparttable}
\end{table}


\renewcommand{\arraystretch}{1.2}
\begin{table}[h!]

\begin{threeparttable}
\caption{ \label{tab:dsphs} Milky way dSphs' properties for \cite{McDaniel:2023bju}}
\centering
\begin{tabular}{ l |r |r| c |c||l |r |r| c |c}
\hline
  Name & $\ell$ & $b$ & Dist. & $\log_{10}({\ensuremath{J_{\textrm{obs}}}})$  & Name & $\ell$ & $b$ & Dist. & $\log_{10}({\ensuremath{J_{\textrm{obs}}}})$\\
   & (deg) & (deg) & (kpc) & {\small ($\log_{10}(\text{GeV}^2 \text{cm}^{-5})$)} & & (deg) & (deg) & (kpc) & {\small ($\log_{10}(\text{GeV}^2 \text{cm}^{-5})$)}\\
  \hline
  Aquarius II & 338.48 & -9.33 & 108.0 & $17.80 \pm 0.55$ &  Bo\"{o}tes II & 209.51 & 12.86 & 42.0 & $18.30 \pm 0.95$\\
Canes Venatici I & 202.01 & 33.55 & 218.0 & $17.42 \pm 0.16$ & Pictor I & 70.95 & -50.29 & 114.0  & $18.00 \pm 0.60$  \\

Carina & 100.41 & -50.96 & 105.0  & $17.83 \pm 0.10$ & Carina II & 114.11 & -58.0 & 36.0 & $18.25 \pm 0.55$  \\

Coma Berenices & 186.75 & 23.91 & 44.0  & $19.00 \pm 0.35$ & Draco & 260.07 & 57.92 & 76.0  & $18.83 \pm 0.12$  \\

Draco II & 238.17 & 64.58 & 22.0 & $18.93 \pm 1.54$ & Eridanus II & 56.09 & -43.53 & 380.0 & $16.60 \pm 0.90$ \\

Fornax & 39.96 & -34.5 & 147.0 & $18.09 \pm 0.10$ & Grus I & 344.18 & -50.18 & 120.0 & $16.50 \pm 0.80$ \\

Hercules & 247.77 & 12.79 & 132.0  & $17.37 \pm 0.53$ & Horologium I & 43.88 & -54.12 & 79.0& $19.00 \pm 0.81$ \\

Hydrus I & 37.39 & -79.31 & 28.0  & $18.33 \pm 0.36^b$ & Leo I & 152.11 & 12.31 & 254.0 & $17.64 \pm 0.13$ \\

Leo II & 168.36 & 22.15 & 233.0 & $17.76 \pm 0.20$ & Leo IV & 173.24 & -0.55 & 154.0& $16.40 \pm 1.08$\\

Leo V & 172.79 & 2.22 & 178.0 & $17.65 \pm 0.97$ & Pegasus III & 336.1 & 5.41 & 215.0 & $18.30 \pm 0.93$  \\

Pisces II & 344.63 & 5.95 & 182.0  & $17.30 \pm 1.04$  & Reticulum II & 53.92 & -54.05 & 30.0 & $18.90 \pm 0.38$ \\

Sagittarius II & 298.16 & -22.07 & 69.0  & $17.35 \pm 1.36^d$ & Segue 1 & 151.75 & 16.08 & 23.0 & $19.12 \pm 0.53$ \\

Sextans & 153.26 & -1.61 & 86.0 & $17.73 \pm 0.12$ & Tucana II & 342.98 & -58.57 & 58.0& $18.97 \pm 0.54$ \\

Tucana IV & 0.73 & -60.85 & 48.0 & $18.40 \pm 0.55^e$ & Ursa Major I & 158.77 & 51.95 & 97.0 & $18.26 \pm 0.28$  \\

Ursa Major II & 132.87 & 63.13 & 32.0 & $19.44 \pm 0.40$ & Ursa Minor & 227.24 & 67.22 & 76.0 & $18.75 \pm 0.12$ \\

Bo\"{o}tes IV & 233.69 & 43.73 & 209.0 & $17.25 \pm 0.60$  & Carina III & 114.63 & -57.9 & 28.0 & $19.70 \pm 0.60$  \\
Centaurus I & 189.59 & -40.9 & 116.0  & $18.14 \pm 0.60$  & Cetus II & 19.47 & -17.42 & 30.0& $19.10 \pm 0.60$  \\

Cetus III & 31.33 & -4.27 & 251.0  & $17.30 \pm 0.60$  & Columba I & 82.86 & -28.01 & 183.0 & $17.60 \pm 0.60$\\

Grus II & 331.02 & -46.44 & 53.0  & $18.40 \pm 0.60$  & Phoenix II & 355.0 & -54.41 & 83.0  & $18.30 \pm 0.60$\\

Canes Venatici II & 194.29 & 34.32 & 160.0  & $17.82 \pm 0.47$ & Pictor II & 101.18 & -59.9 & 46.0 & $18.83 \pm 0.60$ \\

Reticulum III & 56.36 & -60.45 & 92.0  & $18.20 \pm 0.60$ & Tucana V & 354.35 & -63.27 & 55.0  & $18.90 \pm 0.60$ \\

  \hline
\end{tabular}
\label{tableJfactor14}
\end{threeparttable}
\end{table}

\begin{figure}[htbp]
\centering
\includegraphics[width=0.4\textwidth]{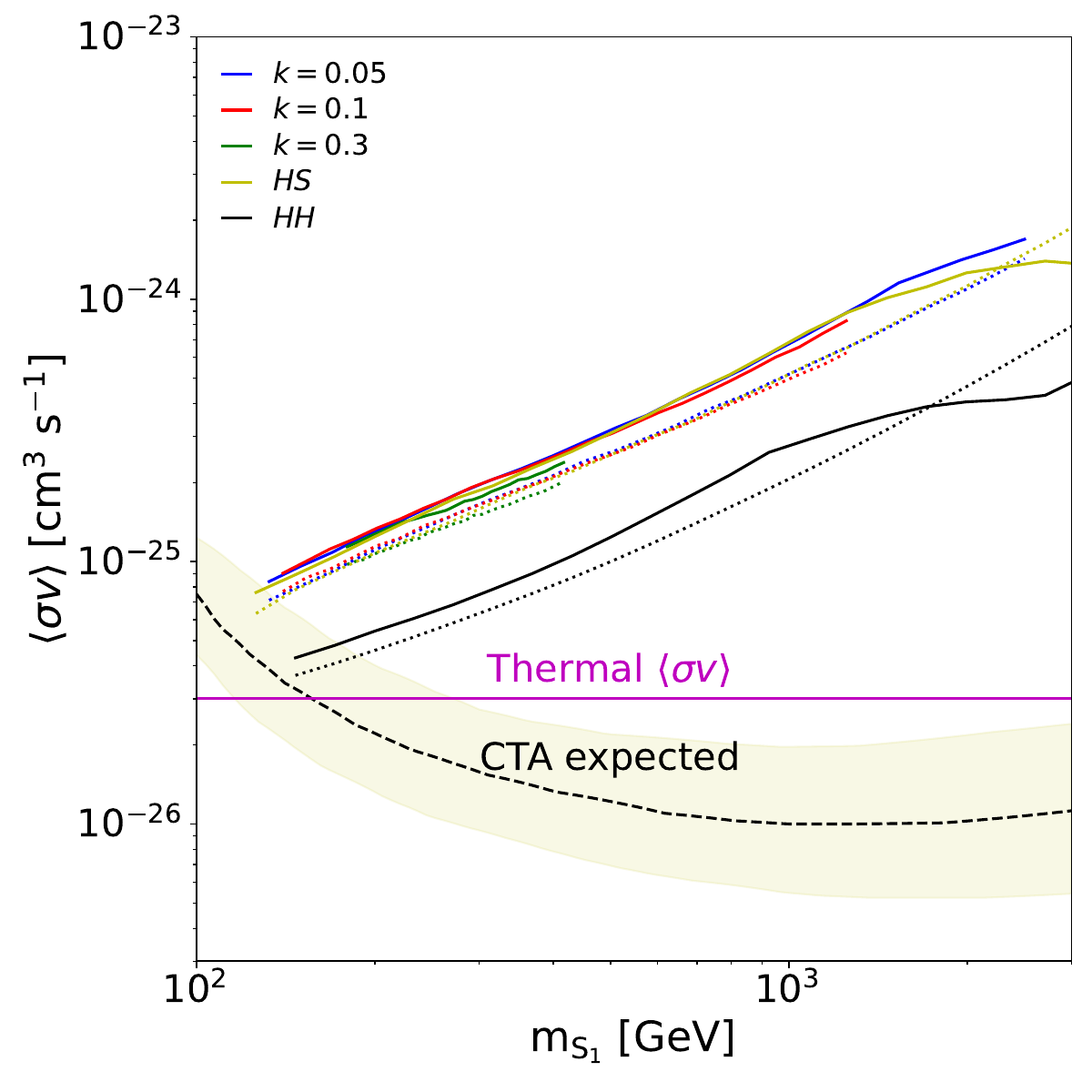}
\includegraphics[width=0.4\textwidth]{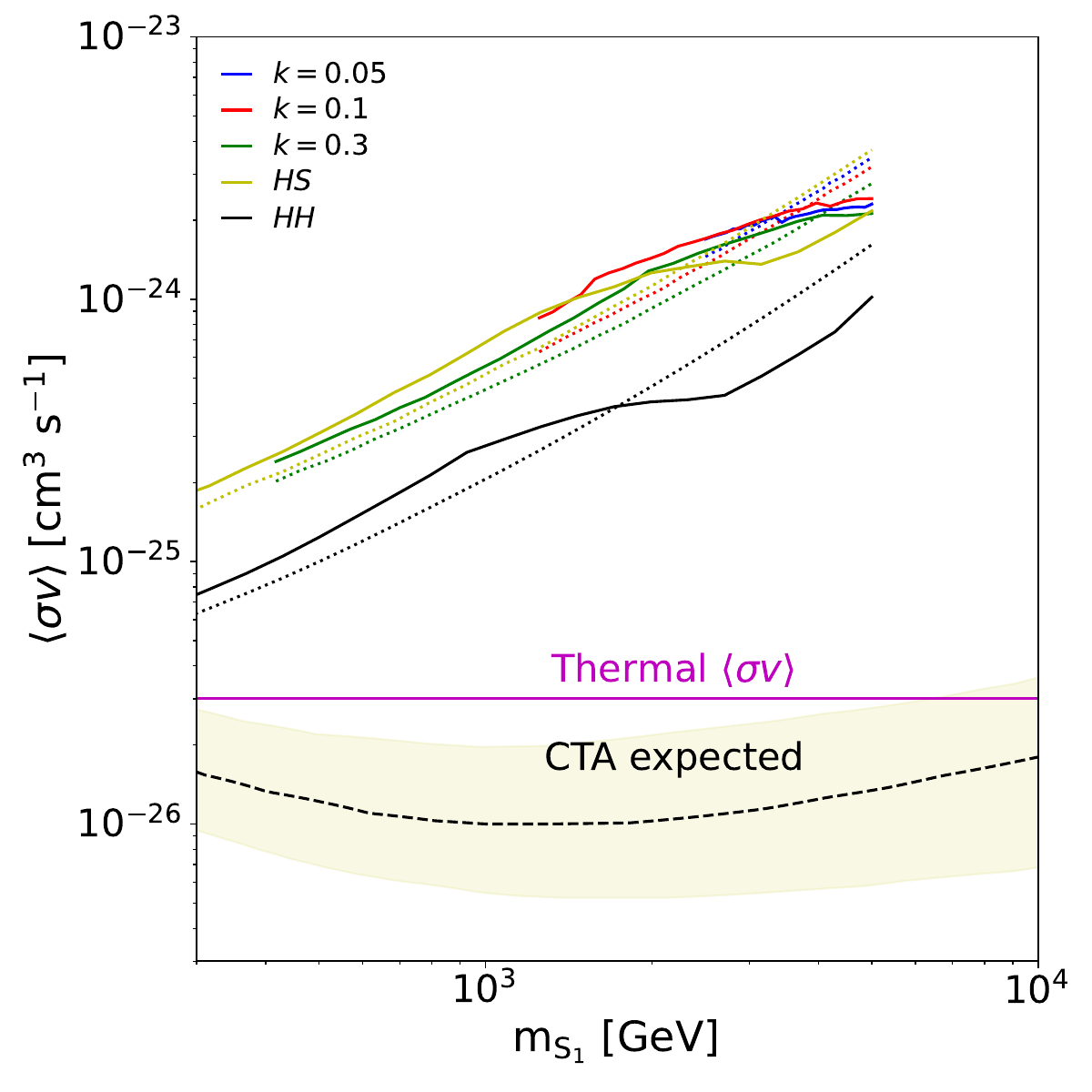}
\caption{Left: Excluding lines with different DM degree of Mass splittings for the $HS_2^*(\to H^{off}+\rm DM)$ mode. Right: Excluding lines with different DM degree of Mass splittings for the $HS_2^*(\to H+\rm DM)$ mode. The 14-year (6-year) data limitations is shown in solid (dotted) lines, and the yellow band is the 1-$\sigma$ uncertainty of CTA expected exclusion sensitivity of $W^+W^-$ annihilation channel.}
\label{ul_result}
\end{figure}
In the Fig.~\ref{ul_result}, using both the 6-year and 14-year Fermi-LAT dSphs data, we demonstrate the upper bounds on the $m_{S_1}-\langle \sigma v\rangle$ plane at a 95\% confidence level (C.L.) with $\rm TS=-2.71$. First of all, the current exclusion is already close to the typical thermal DM cross section  with $\langle \sigma v\rangle\simeq 3\times 10^{-26}\rm cm^{-3}s^{-1}$ for the weak scale DM. But for the relatively heavy DM of sub-TeV or even heavier,  we still need to increase sensitivity orders of magnitude. Next, it is evident that in most of the DM mass region, the 6-year data sets an even tighter constraint than the 14-year data; only in the relatively heavy region of a few TeVs does the latter begin to become more stringent. This is because in the 14-year data, some dSphs, such as Reticulum II, show inconspicuous excesses at the $2-3\sigma$ level~\footnote{It was also noticed in the previous studies~\cite{Fermi-LAT:2015ycq,DiMauro:2021qcf,McDaniel:2023bju}. But there is no single one that meets the detection requirement $\rm TS>25$ adopted by Fermi-LAT, and thus can not be regarded as evidence for DM.}, and therefore a larger $\langle\sigma v\rangle$ is allowed in the 95\% C.L. exclusion. We have to stress that, for simplicity, in this work we did not perform uncertainty analysis associated with the background cosmic rays. If one takes the blank-field analysis, which takes into account both the unknown gamma-ray sources and the uncertainty of backgrounds, then the 14-year data could instead provide the most stringent restrictions~\cite{McDaniel:2023bju}.


The upper bounds in the parameter space with an off-shell Higgs boson signal are shown in the left panel of Fig.~\ref{ul_result}, and we fix the degree of mass splitting $k\equiv \Delta m/m_{S_1}$, taking three typical $k=$0.05 (blue), $0.1$ (red) and $0.3$ (green); the corresponding region of dark matter mass is limited by Eq.~(\ref{window}), and a larger $k$ allows for a narrower interval, with a smaller upper limit. One can see that, for a given DM mass, a larger mass splitting results in a little more stringent bound in most DM mass region, as according to the spectral analysis in the previous section, it produces more photons in the $E_\gamma\lesssim 5$ GeV region; moreover, such photons tend to yield the relatively stronger bound in the likelihood analysis. But overall, the off-shell case behaves quite like the single Higgs boson mode from the semi-annihilation model (gray), which is almost indistinguishable with the $k=0.1$ case, especially for the 6-years data.


The upper bounds on the case with two on-shell Higgs bosons are show on the right panel of Fig.~\ref{ul_result}.  Similarly, a larger mass splitting yields a slightly stronger constraint, except for the DM mass region above $\sim 3$ TeV , where the $k=0.05$ case instead yields a stronger constraint than the $k=0.1$ case (but the $k=0.3$ case still gives the strongest one). Note that it occurs only for the 14-year data, so this is a combined result between spectrum and data. The reason is probably due to that the $S_2^*$ on-shell channel is about to open, and the direct product Higgs is quite soft, and as discussed in the previous section, the $S_2^*$ decay produced Higgs is even softer, so the gamma-ray spectrum produced by these two Higgs boson is soft. 

We end up this section with the comment on comparison with the ${\rm DM}+H$ and $HH$ modes.  As expected, the $HS$ mode confronts with very close exclusion lines to the off-shell mode with a small $k$. But there are still some tiny difference. For instance, a larger $\Delta m\lesssim m_H$ will change the situation, as manifest for the TeV scale DM. For the on-shell mode, the bounds on the ${\rm DM}+H$ mode is weaker with exception in the multi-TeV region. The $HH$ mode is always subject to stricter constraints, and its upper limit of cross section is approximately half of the others.




\subsection{ The heavy dark matter prospect at CTA}

For the TeV-scale heavy dark matter, the limitation of Fermi-LAT to these Higgs bosons in the sky remains relatively weak, because it just observes the gamma-ray energy between 20 MeV and 300 GeV.

In recent years, new indirect detection experiments aiming at probing heavy dark matter are on the table. For instance, the Cherenkov Telescope Array (CTA)~\cite{CTAConsortium:2017dvg}, which has high sensitivity to the photon in the energy range from a few tens of GeV to above 300 TeV, is currently in trial operation. So, it is able to give a strict bound on very heavy DM, and for the $W^+W^-$ channel, the expected exclusion line (with 1$\sigma$ uncertainty) by analysis the Galactic Center data can reach the typical thermal cross section, $\langle\sigma v\rangle\simeq 1\times 10^{-26} \rm cm^3 s^{-1}$.  The specific sensitivity curvtures for the modes we studied need future study, and in this work we take that of the $W^+W^-$ channel as an approximation~\cite{CTA:2020qlo}, and demonstrate it in the plots. It is seen that the complete parameter space of the models studied in this work can be covered, except for the region with substantial coannihilation effect in the DM-companion model.


\section{Conclusion}
\label{sec:conclusion}

The $Z_N$ symmetric DM-companion model offers WIMP DM candidates that may be free of DM direct detection signals, but they still leave indirect detection signals. In this work we study the $Z_3$ symmetric DM-companion model, where the scalar dark matter annihilates into a scalar anti-DM companion and the SM Higgs boson, and the subsequent decay production of the companion could be a DM plus an on-shell (off-shell) Higgs boson. Similar to the conventional semi-annihilation model, our case could evades the stringent constrains from DM direct detection experiments, so we need to search for the photon signal produced by the DM annihilation according to the indirect DM searches.

We study the spectrum produced by our model, and analysis the relationship between the mass splitting $\Delta m$ and the shape of the spectrum, we also make a comparison between our case, Higgs-portal mode and conventional semi-annihilation mode. According to the decay of the DM companion $S_2$, our study is divided into two cases according to the decay of anti-DM companion, the off-shell case and the on-shell case, which is determined by the mass splitting $\Delta m$. We find the photon spectrum in the off-shell case will affected by both mass splitting and DM mass, and it resembles that from the conventional semi-annihilation case. In the on-shell case, however, the spectrum is dominantly affected by the DM mass, and the second Higgs boson from the anti-DM companion decay plays only a subsidiary role in the photon generation.

Finally, we set the upper bound on the $m_{S_1}-\langle\sigma v\rangle$ plane at a $95\%$ C.L level by using both 6-year 14-year Fermi-LAT dSphs data. We find that the 6 year constraint is even tighter than the 14 year one, due to the fact that some inconspicuous excesses at the $2-3\sigma$ level are present in the latter. The $95\%$ CL upper limitation of the $\langle\sigma v\rangle$ in our model reaches about $\simeq 10^{-25}$ $\rm cm^3/s$ at $m_{S_1}\simeq 300$ GeV, which is close to the typical thermal DM cross section, but for the heavy mass region of a few TeV, Fermi-LAT is still not sensitive enough, we expect that the future CTA could cover this region and we will have a detailed study in the future.

We end up the work with a comment on the case that the WIMP DM candidate of the $Z_N$ symmetric DM-companion model is instead a fermion. Then, the resulting indirect detection signals depend on model. A good case in point is provided in Ref.~\cite{Chen:2024arl}, where the companion field further couples to the SM leptonic sector with the help of a doublet $Z_3$ companion, carrying identical SM quantum numbers with the SM Higgs doublet, and then these couplings open decay channel of companion into DM plus neutrino. We can search for such neutrino signal, other than the gamma-ray signal from the direct Higgs boson discussed in this work.

\section*{acknowledgments}
Thanks for the useful discussion with Qiang Yuan. This work was supported by the National Natural Science Foundation of China under Grant No. 12305111 and 11775086.

\end{document}